\documentstyle[aps,prd,eqsecnum]{revtex}

\begin{document}
\title{Master Equation for Quantum Brownian Motion Derived by Stochastic Methods}
\author{Esteban Calzetta}
\address{Departamento de F\'{\i}sica,\\
Universidad de Buenos Aires, Ciudad Universitaria,\\
1428 Buenos Aires, Argentina}
\author{Albert Roura and Enric Verdaguer \thanks{%
Also at Institut de F\'\i sica d'Altes Energies (IFAE), Barcelona, Spain.}}
\address{Departament de F\'{\i}sica Fonamental,\\
Universitat de Barcelona, Av.~Diagonal 647,\\
08028 Barcelona, Spain}
\maketitle

\begin{abstract}
The master equation for a linear open quantum system in a general
environment is derived using a stochastic approach. This is an alternative
derivation to that of Hu, Paz and Zhang, which was based on the direct
computation of path integrals, or to that of Halliwell and Yu, based on the
evolution of the Wigner function for a linear closed quantum system. We
first show by using the influence functinal formalism that the reduced
Wigner function for the open system coincides with a distribution function
resulting from averaging both over the initial conditions and the stochastic
source of a formal Langevin equation. The master equation for the reduced
Wigner function can then be deduced as a Fokker-Planck equation obtained
from the formal Langevin equation.
\end{abstract}


\section{INTRODUCTION}

Open quantum systems are of interest in condensed matter physics \cite
{caldeira83b,leggett87}, quantum optics \cite{walls94}, quantum measurement
theory \cite{zurek81}, nonequilibrium field theory \cite
{calzetta88,calzetta00a,stephens99,calzetta00b}, quantum cosmology \cite
{habib90a,paz91} and semiclassical gravity \cite{hu89,calzetta99}. An open
quantum system consists of a subset of degrees of freedom, whose dynamics
one is interested in, within a larger closed quantum system undergoing
unitary evolution \cite{davies76}. This subsystem of interest is simply
called the ``system'' whereas the remaining degrees of freedom constitute
the ``environment''. In general, the evolution of the system will be
nonunitary and even non-Markovian.

A typical example of an open quantum system is the quantum Brownian motion
(QBM) model, which consists of a single massive particle interacting with an
infinite set of independent harmonic oscillators with a Gaussian initial
state \cite{zwanzig59}. The coupling may be linear both in the system and
environment variables or may be nonlinear in some or all of these variables.
The frequencies of the environment oscillators are distributed according to
a prescribed spectral density function, the simplest case corresponding to
the so-called ohmic environment. Part of the interest of the linear systems
is that they are in many cases exactly solvable and detailed studies of
different aspects of open quantum systems can be performed. One of the
issues that have received much attention in recent years is
environment-induced decoherence as a mechanism to understand the transition
from the quantum to the classical regime \cite{zurek91,zurek93a}.

Certain useful information for an open quantum system is contained in the 
{\em master equation}. The master equation is a differential equation for
the time evolution of the reduced density matrix of an open quantum system.
The master equation for linear coupling and ohmic environment at high
temperature was first deduced by Caldeira and Leggett \cite{caldeira83a}, it
was extended to arbitrary temperature by Unruh and Zurek \cite{unruh89}, and
it was finally obtained for a general environment ({\em i.e.} for an
arbitrary spectral density function) by Hu, Paz and Zhang using path
integrals \cite{hu92}. This result was then extended to the case of
nonlinear coupling by treating the interaction perturbatively up to
quadratic order \cite{hu93}.

The {\em reduced Wigner function} is defined from the reduced density matrix
by an integral transform \cite{wigner32,hillary86}. This function is similar
in many aspects to a distribution function in phase space, although it is
not necessarily positive definite, and the dynamical equation it satisfies
is similar to the {\em Fokker-Planck equation} for classical statistical
systems \cite{wax54,gardiner83,risken89}. This equation is, of course,
entirely equivalent to the master equation for the reduced density matrix
and we will often also refer to it as the master equation. Halliwell and Yu
exploited the fact that the Wigner function for a linear closed quantum
system evolves according to the classical equations of motion to obtain the
equation satisfied by the reduced Wigner function \cite{halliwell96}. The
reduced density matrix has been used to study decoherence induced by the
environment \cite
{joos85,caldeira85,paz93a,zurek93b,giulini96,unruh89,hu92,hu93}. The Wigner
function has also been used in studies of emergence of classicality induced
by an environment \cite{paz93b}, especially in quantum cosmology \cite
{habib90a,paz91}.

Langevin type of equations \cite{zwanzig73,sancho80} as a suitable tool to
study the semiclassical limit have been used recently in semiclassical
gravity and cosmology \cite{calzetta94,hu95,calzetta97a,martin99a,calzetta99}%
. In inflationary cosmology they have been used to describe the stochastic
effect on the inflaton field \cite
{starobinsky86,habib92a,calzetta95,calzetta97c,matacz97,polarski96,kiefer98}
or the stochastic behavior of large-scale gravitational perturbations \cite
{roura99b}, which is important for cosmological structure formation. So far,
in the functional approach the Langevin equation has been mainly restricted
to describe the classical or semiclassical limit. See, however, ref. \cite
{o'connell} for a quantum version of the Langevin equation in operator
language.

A closer look at the influence functional, nevertheless, reveals that a
formal Langevin equation can be extracted from this functional independently
of the existence of a classical limit at least for quadratic influence
actions. This Langevin equation is used to show that the reduced Wigner
function can be written as a formal phase-space distribution function
associated to a stochastic process \cite{calzetta01} (as earlier suggested
in ref. \cite{anglin96}). The master equation governing its time evolution
is then deduced as the corresponding Fokker-Planck equation.

The plan of the paper is the following. In sec. \ref{sec2} we briefly
summarize the essential concepts and results of the influence functional
formalism for linear open quantum systems. In sec. \ref{sec3} we show how
the reduced Wigner function for the system can be expressed as an average
over the different realizations of a stochastic process. This result is used
in sec. \ref{sec4} to give an alternative derivation of the master equation
for a general environment. Finally, we summarize and discuss our results in
sec. \ref{sec5}.

\section{INFLUENCE FUNCTIONAL FORMALISM AND MASTER EQUATION FOR LINEAR OPEN
QUANTUM SYSTEMS}

\label{sec2}

Let us first review a QBM model as an example of linear open quantum system.
We consider a harmonic oscillator of mass $M$, the ``system'', coupled to a
bath of independent harmonic oscillators of mass $m$, the ``environment''.
For simplicity, let us assume that the system and environment are linearly
coupled. The action for the whole set of degrees of freedom is defined by: 
\begin{equation}
S[x,\{q_{j}\}]=S[x]+S[\{q_{j}\}]+S_{int}[x,\{q_{j}\}]\text{,}  \label{2.1}
\end{equation}
where the terms on the right-hand side correspond to the action of the
system, the environment and the interaction term respectively. They are
given by: 
\begin{equation}
S[x]=\int dt(\frac{1}{2}M\dot{x}^{2}-\frac{1}{2}M\Omega ^{2}x^{2})\text{,}
\label{2.2}
\end{equation}
\begin{equation}
S[\{q_{j}\}]=\sum_{j}\int dt(\frac{1}{2}m\dot{q}_{j}^{2}-\frac{1}{2}m\omega
_{j}^{2}q_{j}^{2})\text{,}  \label{2.3}
\end{equation}
\begin{equation}
S_{int}[x,\{q_{j}\}]=\sum_{j}c_{j}\int dtx(t)q_{j}(t)=\int_{0}^{\infty
}d\omega \frac{2m\omega }{\pi c(\omega )}I(\omega )\int dtx(t)q(t;\omega )%
\text{,}  \label{2.4}
\end{equation}
where we introduced the spectral density $I(\omega )=\sum_{j}\pi
c_{j}^{2}(2m\omega _{j})^{-1}\delta (\omega -\omega _{j})$ in the last
equality, $c(\omega )$ and $q(t;\omega )$ are functions such that $c(\omega
_{j})=c_{j}$ and $q(t;\omega _{j})=q_{j}(t)$, $c_{j}$ being
system-environment coupling parameters, and $\Omega $ and $\omega _{j}$ are,
respectively, the system and environment oscillator frequencies. When no
special form is assumed for the spectral density $I(\omega )$, this is
usually referred to as a general environment. One of the most common
particular cases is the so-called Ohmic environment, defined by $I(\omega
)\sim \omega $ (some high frequency cut-off may be sometimes naturally
introduced).

The reduced density matrix for an open quantum system is defined from the
density matrix $\rho $ of the whole system by tracing out the environment
degrees of freedom 
\begin{equation}
\rho _{r}(x_{f},x_{f}^{\prime },t_{f})=\int \prod_{j}dq_{j}\rho
(x_{f},\{q_{j}\},x_{f}^{\prime },\{q_{j}\},t_{f})=\int dx_{i}dx_{i}^{\prime
}J(x_{f},x_{f}^{\prime },t_{f};x_{i},x_{i}^{\prime },t_{i})\rho
_{r}(x_{i},x_{i}^{\prime },t_{i})\text{,}  \label{2.5}
\end{equation}
where the last equation gives the evolution of the reduced density matrix by
means of the propagator $J$, which is defined in a path integral
representation by 
\begin{equation}
J(x_{f},x_{f}^{\prime },t_{f};x_{i},x_{i}^{\prime
},t_{i})=\int\limits_{x(t_{i})=x_{i}}^{x(t_{f})=x_{f}}{\cal D}%
x\int\limits_{x^{\prime }(t_{i})=x_{i}^{\prime }}^{x^{\prime
}(t_{f})=x_{f}^{\prime }}{\cal D}x^{\prime }e^{i(S[x]-S[x^{\prime
}]+S_{IF}[x,x^{\prime }])/\hbar }\text{,}  \label{2.6}
\end{equation}
where $S_{IF}[x,x^{\prime }]$ is the influence action introduced by Feynman
and Vernon \cite{feynman63}. When the system and the environment are
initially uncorrelated, {\em i.e.}, when the initial density matrix
factorizes ($\hat{\rho}(t_{i})=\hat{\rho}_{r}(t_{i})\otimes \hat{\rho}%
_{e}(t_{i})$, where $\hat{\rho}_{r}(t_{i})$ and $\hat{\rho}_{e}(t_{i})$
mean, respectively, the density matrix operators of the system and the
environment at the initial time) the influence functional, defined by $%
F[x,x^{\prime }]=\exp (iS_{IF}[x,x^{\prime }])/\hbar $, can be expressed in
the following way: 
\begin{eqnarray}
F[x,x^{\prime }] &=&\prod_{j}\int dq_{j}^{(f)}dq_{j}^{(i)}dq_{j}^{\prime
(i)}\int\limits_{q_{j}(t_{i})=q_{j}^{(i)}}^{q_{j}(t_{f})=q_{j}^{(f)}}{\cal D}%
q_{j}\int\limits_{q_{j}^{\prime }(t_{i})=q_{j}^{\prime (i)}}^{q_{j}^{\prime
}(t_{f})=q_{j}^{(f)}}{\cal D}q_{j}^{\prime }\exp \left[ \frac{i}{\hbar }%
\left( S[\{q_{j}\}]-S[\{q_{j}^{\prime }\}]+S[x,\{q_{j}\}]\right. \right. 
\nonumber \\
&&\left. \left. -S[x^{\prime },\{q_{j}^{\prime }\}]\right) \right] \cdot
\rho _{e}(\{q_{j}^{(i)}\},\{q_{j}^{\prime (i)}\},t_{i})\text{.}  \label{2.7}
\end{eqnarray}
When the initial density matrix for the environment $\rho
_{e}(\{q_{j}^{(i)}\},\{q_{j}^{\prime (i)}\},t_{i})$ is Gaussian, the path
integrals can be exactly performed and one obtains \cite
{feynman63,caldeira83a}: 
\begin{equation}
S_{IF}[x,x^{\prime }]=-2\int_{t_{i}}^{t_{f}}ds\int_{t_{i}}^{s}ds^{\prime
}\Delta (s)D(s,s^{\prime })X(s^{\prime })+\frac{i}{2}\int_{t_{i}}^{t_{f}}ds%
\int_{t_{i}}^{t_{f}}ds^{\prime }\Delta (s)N(s,s^{\prime })\Delta (s^{\prime
})\text{,}  \label{2.8}
\end{equation}
where $X(s)\equiv (x(s)+x^{\prime }(s))/2$ and $\Delta (s)\equiv x^{\prime
}(s)-x(s)$. The kernels $D(s,s^{\prime })$ and $N(s,s^{\prime })$ are called
the dissipation and noise kernel, respectively.

For environments consisting of an infinite number of oscillators it is
especially convenient to rewrite the first term on the right-hand side of
Eq. (\ref{2.8}) as 
\begin{equation}
\int_{t_{i}}^{t_{f}}ds\int_{t_{i}}^{t_{f}}ds^{\prime }\Delta
(s)H_{bare}(s,s^{\prime })X(s^{\prime })\text{,}  \label{3.1}
\end{equation}
where we defined $H_{bare}(s,s^{\prime })$ as formally equivalent to $%
-2D(s,s^{\prime })\theta (s-s^{\prime })$. Being the product of two
distributions the latter expression is not well defined in general and
suitable regularization and renormalization may be required; see \cite
{roura99a} for details. The local divergences present in $%
H_{bare}(s,s^{\prime })=H(s,s^{\prime })+H_{div}\delta (s-s^{\prime })$ can
be canceled by suitable counterterms $\Omega _{div}$ in the bare frequency
of the system $\Omega =\Omega _{ren}+\Omega _{div}$. From now on we will
consider that this infinite renormalization, if necessary, has already been
performed so that both $\Omega _{ren}$ and $H(s,s^{\prime })$ are free of
divergences.

From Eqs. (\ref{2.5}) and (\ref{2.6}) a differential equation for the
system's reduced density matrix known as the master equation can be derived.
The expression for a general environment was first obtained by Hu, Paz and
Zhang using a path integral approach \cite{hu92} (see \cite{paz94} for a
slightly different derivation):

\begin{eqnarray}
i\hbar \frac{\partial \rho _{r}}{\partial t} &=&-\frac{\hbar ^{2}}{2M}\left( 
\frac{\partial ^{2}}{\partial x^{2}}-\frac{\partial ^{2}}{\partial x^{\prime
2}}\right) \rho _{r}+\frac{1}{2}M\Omega ^{2}(x^{2}-x^{\prime 2})\rho _{r}+%
\frac{1}{2}M\delta \Omega ^{2}(t)(x^{2}-x^{\prime 2})\rho _{r}  \nonumber \\
&&-i\hbar A(t)(x-x^{\prime })\left( \frac{\partial }{\partial x}-\frac{%
\partial }{\partial x^{\prime }}\right) \rho _{r}+\hbar B(t)(x-x^{\prime
})\left( \frac{\partial }{\partial x}+\frac{\partial }{\partial x^{\prime }}%
\right) \rho _{r}-iMC(t)(x-x^{\prime })^{2}\rho _{r}\text{,}  \label{2.12}
\end{eqnarray}
where the functions $\delta \Omega ^{2}(t)$, $A(t)$, $B(t)$ and $C(t)$
represent a frequency shift, a dissipation factor and two diffusive factors,
respectively. For explicit expressions of these functions see sec. \ref{sec4}%
. An alternative representation for the system reduced density matrix is the
reduced Wigner function $W_{r}(X,p,t)$ defined as 
\begin{equation}
W_{r}(X,p,t)=\frac{1}{2\pi \hbar }\int_{-\infty }^{\infty }d\Delta
e^{ip\Delta /\hbar }\rho _{r}(X-\Delta /2,X+\Delta /2,t)\text{.}
\label{2.13}
\end{equation}
It follows immediately that the master equation (\ref{2.12}) can be written
in the following equivalent form: 
\begin{equation}
\frac{\partial W_{r}}{\partial t}=\{H_{R},W_{r}\}_{PB}+2A(t)\frac{\partial
(pW_{r})}{\partial p}+\hbar B(t)\frac{\partial ^{2}W_{r}}{\partial q\partial
p}+\hbar MC(t)\frac{\partial ^{2}W_{r}}{\partial p^{2}}\text{,}  \label{2.14}
\end{equation}
where $\{H_{R},W_{r}\}_{PB}\equiv -(p/M)\partial W_{r}/\partial q+M\Omega
_{R}^{2}(t)q\partial W_{r}/\partial p$ with $\Omega _{R}^{2}(t)=\Omega
^{2}+\delta \Omega ^{2}(t)$. This equation was directly derived by Halliwell
and Yu \cite{halliwell96} exploiting the fact that the Wigner function for
the whole closed quantum system evolves according to the classical equations
of motion. Note that Eq. (\ref{2.14}) is formally similar to the
Fokker-Planck equation for a distribution function.

\section{STOCHASTIC FORMAL EXPRESSION FOR THE REDUCED WIGNER FUNCTION}

\label{sec3}

In this section we show that the reduced Wigner function can be written as a
formal distribution function for some stochastic process (see ref. \cite
{calzetta01} for a detailed exposition). This will be the key starting point
in the derivation of the master equation given in the next section.

In order to find an explicit expression for the reduced density matrix (\ref
{2.5}) at a time $t_{f}$, we need to compute the path integrals appearing in
Eq. (\ref{2.6}) for the reduced density matrix propagator. From now on we
will consider $\hbar =1$. After integrating the system action by parts and
performing the Gaussian path integral for $\Delta (t)$ with $\Delta _{i}$
and $\Delta _{f\text{ }}$ fixed, we obtain

\begin{equation}
\int_{X_{i}}^{X_{f}}{\cal D}X\int_{\Delta _{i}}^{\Delta _{f}}{\cal D}\Delta
e^{i\Delta \cdot L\cdot X}e^{-\frac{1}{2}\Delta \cdot N\cdot \Delta }=\left(
\det \frac{N}{2\pi }\right) ^{-\frac{1}{2}}\int_{X_{i}}^{X_{f}}{\cal D}Xe^{-%
\frac{1}{2}(L\cdot X)\cdot N^{-1}\cdot (L\cdot X)}\text{,}  \label{3.3}
\end{equation}
where $L(t,t^{\prime })\equiv M\left( \frac{d^{2}}{dt^{\prime 2}}+\Omega
_{ren}^{2}\right) \delta (t-t^{\prime })+H(t,t^{\prime })$. Taking into
account the surface terms arising from the integration by parts of the
system action and definition (\ref{2.13}) for the reduced Wigner function,
the result of the integration over $\Delta _{i}$ gives 
\begin{equation}
\rho _{r}(X_{f}-\Delta _{f}/2,X_{f}+\Delta _{f}/2,t_{f})=2\pi \left( \det 
\frac{N}{2\pi }\right) ^{-\frac{1}{2}}\int_{-\infty }^{\infty
}dX_{i}\int_{X_{i}}^{X_{f}}{\cal D}Xe^{-\frac{1}{2}(L\cdot X)\cdot
N^{-1}\cdot (L\cdot X)}e^{-iM\dot{X}_{f}\Delta _{f}}W_{r}(X_{i},M\dot{X}%
_{i},t_{i})\text{.}  \label{3.4}
\end{equation}

The next step to perform is the following functional change: 
\begin{equation}
X(t)\longrightarrow \left\{ X_{i}=X(t_{i})\text{, }p_{i}\equiv M\dot{X}_{i}=M%
\dot{X}(t_{i})\text{, }\xi (t)=(L\cdot X)(t)\text{ }\right\} \text{.}
\label{3.5}
\end{equation}
Note that with this change the function $X(t)$ gets substituted by the
initial conditions $(X_{i},p_{i})$ and the function $\xi (t)$ in the path
integration. It is important to note that at this point the function $\xi
(t) $ is not a stochastic process but just a function over which a path
integral is performed. The functional change (\ref{3.5}) is invertible as
can be explicitly seen: 
\begin{equation}
\left\{ X_{i}\text{, }p_{i}\text{, }\xi (t)\right\} \longrightarrow
X(t)=X_{o}(t)+\int_{t_{i}}^{t}dt^{\prime }G_{ret}(t,t^{\prime })\xi
(t^{\prime })\text{,}  \label{3.6}
\end{equation}
where $G_{ret}(t^{\prime },t^{\prime \prime })$ is the retarded ({\em i.e.}, 
$G_{ret}(t^{\prime },t^{\prime \prime })=0$ for $t^{\prime }\leq t^{\prime
\prime }$) Green function for the linear integro-differential operator
associated to the kernel $L(t,t^{\prime })$, and $X_{inh}(t)=$ $%
\int_{t_{i}}^{t}dt^{\prime }G_{ret}(t,t^{\prime })\xi (t^{\prime })$ is a
solution of the inhomogeneous equation $(L\cdot X_{inh})(t)=\xi (t)$ with
initial conditions $X_{inh}(t_{i})=0$ and $\left. \partial X_{inh}(t^{\prime
})/\partial t^{\prime }\right| _{t^{\prime }=t_{i}}=0$. On the other hand, $%
X_{o}(t)$ is a solution of the homogeneous equation $(L\cdot X_{o})(t)=0$,
with initial conditions $X_{o}(t_{i})=X_{i}$ and $\dot{X}(t_{i})=p_{i}/M$.
Since the change is linear, the Jacobian functional determinant will be a
constant (this can be clearly seen by skeletonizing the path integral).
After performing the functional change, we obtain 
\begin{equation}
\rho _{r}(X_{f}-\Delta _{f}/2,X_{f}+\Delta _{f}/2,t_{f})=K\int_{-\infty
}^{\infty }dX_{i}\int_{-\infty }^{\infty }dp_{i}\int {\cal D}\xi \delta
(X(t_{f})-X_{f})e^{-\frac{1}{2}\xi \cdot N^{-1}\cdot \xi }e^{-iM\dot{X}%
(t_{f})\Delta _{f}}W_{r}(X_{i},p_{i},t_{i})\text{,}  \label{3.7}
\end{equation}
where the delta function $\delta (X(t_{f})-X_{f})$ was introduced to
restrict the functional integral $\int {\cal D}\xi $ with free ends, in
order to take into account the restriction on the final points of the
allowed paths for the integral $\int^{X_{f}}{\cal D}X$ appearing in Eq. (\ref
{3.4}). The contribution from the Jacobian has been included in the constant 
$K$. By demanding the reduced density matrix to be normalized, {\em i.e.},
that $Tr\rho _{r}(t_{f})=1$, provided that the initial Wigner function is
properly normalized, this constant can be determined to be 
\begin{equation}
K=\left[ \int {\cal D}\xi e^{-\frac{1}{2}\xi \cdot N^{-1}\cdot \xi }\right]
^{-1}=\left[ \det (2\pi N)\right] ^{-\frac{1}{2}}\text{.}  \label{3.9}
\end{equation}
Finally, using the definition (\ref{2.13}) for the Wigner function and the
fact that $(2\pi )^{-1}\int_{-\infty }^{\infty }d\Delta _{f}e^{ip_{f}\Delta
_{f}}e^{-iM\dot{X}(t_{f})\Delta _{f}}=\delta (M\dot{X}(t_{f})-p_{f})$, we
get an expression for the reduced Wigner function 
\begin{equation}
W_{r}(X_{f},p_{f},t_{f})=K\int_{-\infty }^{\infty }dX_{i}\int_{-\infty
}^{\infty }dp_{i}\int {\cal D}\xi \delta (X(t_{f})-X_{f})\delta (M\dot{X}%
(t_{f})-p_{f})e^{-\frac{1}{2}\xi \cdot N^{-1}\cdot \xi
}W_{r}(X_{i},p_{i},t_{i})\text{,}  \label{3.11}
\end{equation}
which can be written in the following suggestive way: 
\begin{equation}
W_{r}(X_{f},p_{f},t_{f})=\left\langle \left\langle \delta
(X(t_{f})-X_{f})\delta (M\dot{X}(t_{f})-p_{f})\right\rangle _{\xi
}\right\rangle _{X_{i},p_{i}}\text{,}  \label{3.12}
\end{equation}
where 
\begin{eqnarray}
\left\langle ...\right\rangle _{\xi } &\equiv &\left[ \det (2\pi N)\right]
^{-\frac{1}{2}}\int {\cal D}\xi ...e^{-\frac{1}{2}\xi \cdot N^{-1}\cdot \xi }%
\text{,}  \label{3.13} \\
\left\langle ...\right\rangle _{X_{i},p_{i}} &\equiv &\int_{-\infty
}^{\infty }dX_{i}\int_{-\infty }^{\infty }dp_{i}...W_{r}(X_{i},p_{i},t_{i})%
\text{.}  \label{3.14}
\end{eqnarray}
Thus the reduced Wigner function can be interpreted as an average over a
Gaussian stochastic process $\xi (t)$ with $\left\langle \xi
(t)\right\rangle _{\xi }=0$ and $\left\langle \xi (t)\xi (t^{\prime
})\right\rangle _{\xi }=N(t,t^{\prime })$ as well as an average over the
initial conditions characterized by a distribution function $%
W_{r}(X_{i},p_{i},t_{i})$. It is only after formally interpreting $\xi (t)$
as a stochastic process characterized by Eq. (\ref{3.13}) that the equation
defining $\xi (t)$ in the functional change (\ref{3.5}) 
\begin{equation}
(L\cdot X)(t)=\xi (t)\text{,}  \label{3.14b}
\end{equation}
can be regarded as a Langevin equation. We insist that, in general, Eq. (\ref
{3.14b}) is not meant to describe the actual trajectories of the system, but
it should rather be regarded as a formal tool. We should also remark that $%
X(t_{f})$ and $\dot{X}(t_{f})$ in Eq. (\ref{3.12}) correspond to solutions
of the Langevin equation (\ref{3.14b}) for a given realization of $\xi (t)$,
and that $X_{f}$ and $p_{f}$ are coordinates of a point in phase space.

Note, in addition, that although $W_{r}(X_{i},p_{i},t_{i})$ is real, which
follows from the hermiticity of the density matrix, and properly normalized,
in general it is not positive everywhere and, thus, cannot be considered as
a probability distribution. The fact that the Wigner function cannot be
interpreted as a phase space probability density is crucial since most of
the nonclassical features of the quantum state are tightly related to the
Wigner function having negative values. For instance, a coherent
superposition state is typically characterized by the Wigner function
presenting strong oscillations with negative values in the minima \cite
{paz93b,giulini96}, which are closely connected to interference terms.

\section{FROM LANGEVIN TO FOKKER-PLANCK: DERIVATION OF THE MASTER EQUATION}

\label{sec4}

As mentioned above there is a simple one-to-one correspondence between any
density matrix and the associated Wigner function introduced in (\ref{2.13}%
). Taking this correspondence into account, the equation satisfied by the
reduced Wigner function is equivalent to the master equation satisfied by
the reduced density matrix.

Equation (\ref{3.12}) shows that the reduced Wigner function can be
interpreted as a formal distribution in phase space. By deriving it with
respect to time and using the Langevin-type equation in (\ref{3.14b}), one
can obtain a Fokker-Planck differential equation describing the time
evolution of the system's reduced Wigner function.

The derivation of the Fokker-Planck equation from the Langevin equation with
local dissipation is well understood (see ref. \cite{sancho80}). However, in
our case the existence of nonlocal dissipation makes it convenient to review
the main steps. Let us begin by computing $\partial W_{r}/\partial t$ from
expression (\ref{3.12}), 
\begin{eqnarray}
\frac{\partial W_{r}(X,p,t)}{\partial t} &=&\left\langle \left\langle \dot{X}%
(t)\delta ^{\prime }(X(t)-X)\delta (M\dot{X}(t)-p)\right\rangle _{\xi
}\right\rangle _{X_{i},p_{i}}+\left\langle \left\langle \delta (X(t)-X)M%
\ddot{X}(t)\delta ^{\prime }(M\dot{X}(t)-p)\right\rangle _{\xi
}\right\rangle _{X_{i},p_{i}}  \nonumber \\
&=&-\frac{p}{M}\frac{\partial W_{r}(X,p,t)}{\partial X}-\frac{\partial }{%
\partial p}\left\langle \left\langle \delta (X(t)-X)M\ddot{X}(t)\delta (M%
\dot{X}(t)-p)\right\rangle _{\xi }\right\rangle _{X_{i},p_{i}}\text{,}
\label{3.15}
\end{eqnarray}
where the fact that $\dot{X}(t)$, $\partial /\partial X(t)$ and $\partial
/\partial \dot{X}(t)$ may be replaced by $p/M$, $-\partial /\partial X$ and $%
-\partial /\partial p$ respectively, since they are multiplying the delta
functions, was used in the second equality. Let us now concentrate on the
expectation value appearing in the last term and recall the expectation
values defined in (\ref{3.13})-(\ref{3.14}). We will consider the
Langevin-type equation 
\begin{equation}
(L\cdot X)(t^{\prime })=\xi (t^{\prime })\text{,}  \label{3.16}
\end{equation}
corresponding to the functional change (\ref{3.5}) and substitute the
corresponding expression for $M\ddot{X}(t)$ so that the last expectation
value in (\ref{3.15}) can be written as 
\begin{equation}
-M\Omega _{ren}^{2}XW_{r}(X,p,t)+\left\langle \left\langle \left(
-\int_{t_{i}}^{t}dtH(t,t^{\prime })X(t^{\prime })+\xi (t)\right) \delta
(X(t)-X)\delta (M\dot{X}(t)-p)\right\rangle _{\xi }\right\rangle
_{X_{i},p_{i}}\text{.}  \label{3.17}
\end{equation}
Any solution of Eq. (\ref{3.16}) can be written as 
\begin{equation}
X(t^{\prime })=X_{h}(t^{\prime })+\int_{t^{\prime }}^{t}dt^{\prime \prime }%
\tilde{G}_{adv}(t^{\prime },t^{\prime \prime })\xi (t^{\prime \prime })\text{%
,}  \label{3.18}
\end{equation}
where $X_{h}(t^{\prime })$ is a solution of the homogeneous equation $%
(L\cdot X)(t^{\prime })=0$ such that $X_{h}(t)=X$, $\dot{X}_{h}(t)=p/M$ and $%
\tilde{G}_{adv}(t^{\prime },t^{\prime \prime })$ is the advanced ({\em i.e.}%
, $\tilde{G}_{adv}(t^{\prime },t^{\prime \prime })=0$ for $t^{\prime }\geq
t^{\prime \prime }$) Green function for the linear integro-differential
operator associated to the kernel $L(t,t^{\prime })$. The particular
solution of the inhomogeneous Eq. (\ref{3.16}) 
\begin{equation}
\tilde{X}_{inh}(t^{\prime })=\int_{t^{\prime }}^{t}dt^{\prime \prime }\tilde{%
G}_{adv}(t^{\prime },t^{\prime \prime })\xi (t^{\prime \prime })
\end{equation}
has boundary conditions $\tilde{X}_{inh}(t)=0$, $\left. \partial \tilde{X}%
_{inh}(t^{\prime })/\partial t^{\prime }\right| _{t^{\prime }=t}=0$. Both $%
X_{h}(t^{\prime })$ and $\tilde{G}_{adv}(t^{\prime },t^{\prime \prime })$
can be expressed in terms of the homogeneous solutions $u_{1}(t^{\prime })$
and $u_{2}(t^{\prime })$, which satisfy $u_{1}(t_{i})=1$, $u_{1}(t)=0$ and $%
u_{2}(t_{i})=0$, $u_{2}(t)=1$ respectively: 
\begin{eqnarray}
X_{h}(t^{\prime }) &=&X\left( u_{2}(t^{\prime })-\frac{\dot{u}_{2}(t)}{\dot{u%
}_{1}(t)}u_{1}(t^{\prime })\right) +\frac{(p/M)}{\dot{u}_{1}(t)}%
u_{1}(t^{\prime })\text{,}  \label{3.19} \\
\tilde{G}_{adv}(t^{\prime },t^{\prime \prime }) &=&-\frac{1}{M}\frac{%
u_{1}(t^{\prime })u_{2}(t^{\prime \prime })-u_{2}(t^{\prime
})u_{1}(t^{\prime \prime })}{\dot{u}_{1}(t^{\prime \prime })u_{2}(t^{\prime
\prime })-\dot{u}_{2}(t^{\prime \prime })u_{1}(t^{\prime \prime })}\theta
(t^{\prime \prime }-t^{\prime })\text{.}  \label{3.20}
\end{eqnarray}
We use the advanced propagator so that there is no dependence on the initial
conditions at time $t^{\prime }=t_{i}$ coming from the homogeneous solution
but just on the final conditions at time $t^{\prime }=t$, {\em i.e.}, on
those the Fokker-Planck equation is written in terms of. Using expression (%
\ref{3.18}) the first term within the expectation value appearing in Eq. (%
\ref{3.17}) can be reexpressed as 
\begin{eqnarray}
&&\int_{t_{i}}^{t}dtH(t,t^{\prime })\left\langle \left\langle X(t^{\prime
})\delta (X(t)-X)\delta (M\dot{X}(t)-p)\right\rangle _{\xi }\right\rangle
_{X_{i},p_{i}}  \nonumber \\
&=&\int_{t_{i}}^{t}dt^{\prime }H(t,t^{\prime })X_{h}(t^{\prime
})W_{r}(X,p,t)+\int_{t_{i}}^{t}dt^{\prime }\int_{t^{\prime }}^{t}dt^{\prime
\prime }H(t,t^{\prime })\tilde{G}_{adv}(t^{\prime },t^{\prime \prime
})\left\langle \left\langle \xi (t^{\prime \prime })\delta (X(t)-X)\delta (M%
\dot{X}(t)-p)\right\rangle _{\xi }\right\rangle _{X_{i},p_{i}}\text{.}
\label{3.21}
\end{eqnarray}
The first term on the right-hand side can in turn be written as 
\begin{equation}
-\left( M\delta \Omega (t)X+2A(t)p\right) W_{r}(X,p,t)\text{,}  \label{3.22}
\end{equation}
where 
\begin{eqnarray}
\delta \Omega (t) &=&\frac{1}{M}\int_{t_{i}}^{t}dt^{\prime }H(t,t^{\prime
})[u_{2}(t^{\prime })-(\dot{u}_{2}(t)/\dot{u}_{1}(t))u_{1}(t^{\prime })]%
\text{,}  \label{3.23} \\
A(t) &=&\frac{1}{2}(M\dot{u}_{1}(t))^{-1}\int_{t_{i}}^{t}dt^{\prime
}H(t,t^{\prime })u_{1}(t^{\prime })\text{.}  \label{3.24}
\end{eqnarray}

In order to find an expression for $\left\langle \xi (t^{\prime })\delta
(X(t)-X)\delta (M\dot{X}(t)-p)\right\rangle _{\xi }$ we use Novikov's
formula for Gaussian stochastic processes \cite{novikov65}, which
corresponds essentially to use (\ref{3.13}) and functionally integrate by
parts with respect to $\xi (t)$, 
\begin{equation}
\left\langle \xi (t^{\prime })F(t;\xi ]\right\rangle _{\xi
}=\int_{t_{i}}^{t}dt^{\prime \prime }N(t^{\prime },t^{\prime \prime
})\left\langle \delta F(t;\xi ]/\delta \xi (t^{\prime \prime })\right\rangle
_{\xi }\text{.}  \label{3.25}
\end{equation}
We then obtain the following expression: 
\begin{eqnarray}
\left\langle \xi (t^{\prime })\delta (X(t)-X)\delta (M\dot{X}%
(t)-p)\right\rangle _{\xi } &=&\int_{t_{i}}^{t}dt^{\prime \prime \prime
}\int_{t_{i}}^{t}dt^{\prime \prime }N(t^{\prime },t^{\prime \prime
})\left\langle \left( \frac{\delta X(t^{\prime \prime \prime })}{\delta \xi
(t^{\prime \prime })}\frac{\delta }{\delta X(t^{\prime \prime \prime })}%
\right. \right.  \nonumber \\
&&+\left. \left. \frac{\delta \dot{X}(t^{\prime \prime \prime })}{\delta \xi
(t^{\prime \prime })}\frac{\delta }{\delta \dot{X}(t^{\prime \prime \prime })%
}\right) \delta (X(t)-X)\delta (M\dot{X}(t)-p)\right\rangle _{\xi } 
\nonumber \\
&=&\int_{t_{i}}^{t}dt^{\prime \prime \prime }\int_{t_{i}}^{t}dt^{\prime
\prime }N(t^{\prime },t^{\prime \prime })\delta (t^{\prime \prime \prime
}-t)\left\langle -\left( \frac{\delta X(t^{\prime \prime \prime })}{\delta
\xi (t^{\prime \prime })}\frac{\partial }{\partial X}+M\frac{\delta \dot{X}%
(t^{\prime \prime \prime })}{\delta \xi (t^{\prime \prime })}\frac{\partial 
}{\partial p}\right) \right.  \nonumber \\
&&\left. \delta (X(t)-X)\delta (M\dot{X}(t)-p)\right\rangle _{\xi } 
\nonumber \\
&=&\int_{t_{i}}^{t}dt^{\prime \prime }N(t^{\prime },t^{\prime \prime
})\left\langle -\left( \frac{\delta X(t)}{\delta \xi (t^{\prime \prime })}%
\frac{\partial }{\partial X}+M\frac{\delta \dot{X}(t)}{\delta \xi (t^{\prime
\prime })}\frac{\partial }{\partial p}\right) \right.  \nonumber \\
&&\left. \delta (X(t)-X)\delta (M\dot{X}(t)-p)_{\xi }\right\rangle \text{,}
\label{3.26}
\end{eqnarray}
where we used again the presence of the delta functions to substitute the
functional derivatives $\delta /\delta X(t^{\prime \prime \prime })$ and $%
\delta /\delta \dot{X}(t^{\prime \prime \prime })$ by $-\delta (t^{\prime
\prime \prime }-t)\cdot \partial /\partial X$ and $-\delta (t^{\prime \prime
\prime }-t)\cdot M\cdot \partial /\partial p$, respectively, in the second
equality. Functionally differentiating with respect to $\xi (t^{\prime
\prime })$ expression (\ref{3.6}) for $X(t)$ and analogously for $\dot{X}(t)$
we get 
\begin{mathletters}
\begin{eqnarray}
\frac{\delta X(t^{\prime })}{\delta \xi (t^{\prime \prime })}
&=&G_{ret}(t^{\prime },t^{\prime \prime })\text{,}  \label{3.27} \\
\frac{\delta \dot{X}(t^{\prime })}{\delta \xi (t^{\prime \prime })} &=&\frac{%
\partial }{\partial t^{\prime }}G_{ret}(t^{\prime },t^{\prime \prime })\text{%
,}  \label{3.28}
\end{eqnarray}
which after substitution into (\ref{3.26}) leads to 
\end{mathletters}
\begin{equation}
\left\langle \xi (t^{\prime })\delta (X(t)-X)\delta (M\dot{X}%
(t)-p)\right\rangle _{\xi }=-\int_{t_{i}}^{t}dt^{\prime \prime }N(t^{\prime
},t^{\prime \prime })\left( G_{ret}(t,t^{\prime \prime })\frac{\partial }{%
\partial X}+M\frac{\partial G_{ret}(t,t^{\prime \prime })}{\partial
t^{\prime }}\frac{\partial }{\partial p}\right) W_{r}(X,p,t)\text{.}
\label{3.29}
\end{equation}
The retarded Green function can also be expressed in terms of the solutions
of the homogeneous equation $u_{1}(t)$ and $u_{2}(t)$, which were previously
introduced, as 
\begin{equation}
G_{ret}(t^{\prime },t^{\prime \prime })=\frac{1}{M}\frac{u_{1}(t^{\prime
})u_{2}(t^{\prime \prime })-u_{2}(t^{\prime })u_{1}(t^{\prime \prime })}{%
\dot{u}_{1}(t^{\prime \prime })u_{2}(t^{\prime \prime })-\dot{u}%
_{2}(t^{\prime \prime })u_{1}(t^{\prime \prime })}\theta (t^{\prime
}-t^{\prime \prime })\text{.}  \label{3.30}
\end{equation}
Note that it is important to use now the expression in terms of the retarded
propagator $G_{ret}$ and the initial conditions $X_{i}$ and $p_{i}$ (at time 
$t^{\prime }=t_{i}$), since the ``final'' conditions $X(t)$ and $M\dot{X}(t)$
depend on $\xi (t^{\prime \prime })$ (for $t^{\prime \prime }<t$). Putting
all the terms together, {\em i.e.}, (\ref{3.17}), (\ref{3.21}) and (\ref
{3.29}), we reach the final expression for (\ref{3.15}): 
\begin{equation}
\frac{\partial W_{r}}{\partial t}=\{H_{R},W_{r}\}_{PB}+2A(t)\frac{\partial
(pW_{r})}{\partial p}+B(t)\frac{\partial ^{2}W_{r}}{\partial X\partial p}%
+MC(t)\frac{\partial ^{2}W_{r}}{\partial p^{2}}\text{,}  \label{3.31}
\end{equation}
where $\delta \Omega (t)$ and $A(t)$ are given by Eqs. (\ref{3.23}) and (\ref
{3.24}), and 
\begin{eqnarray}
B(t) &=&\int_{t_{i}}^{t}dt^{\prime \prime \prime }N(t,t^{\prime \prime
\prime })G_{ret}(t,t^{\prime \prime \prime })-\int_{t_{i}}^{t}dt^{\prime
}H(t,t^{\prime })\int_{t^{\prime }}^{t}dt^{\prime \prime }\tilde{G}%
_{adv}(t^{\prime },t^{\prime \prime })\int_{t_{i}}^{t}dt^{\prime \prime
\prime }N(t^{\prime \prime },t^{\prime \prime \prime })G_{ret}(t,t^{\prime
\prime \prime })\text{,}  \label{3.32} \\
C(t) &=&\int_{t_{i}}^{t}dt^{\prime \prime \prime }N(t,t^{\prime \prime
\prime })\frac{\partial G_{ret}(t,t^{\prime \prime \prime })}{\partial t}%
-\int_{t_{i}}^{t}dt^{\prime }H(t,t^{\prime })\int_{t^{\prime
}}^{t}dt^{\prime \prime }\tilde{G}_{adv}(t^{\prime },t^{\prime \prime
})\int_{t_{i}}^{t}dt^{\prime \prime \prime }N(t^{\prime \prime },t^{\prime
\prime \prime })\frac{\partial G_{ret}(t,t^{\prime \prime \prime })}{%
\partial t^{\prime \prime }}\text{.}  \label{3.33}
\end{eqnarray}
The last two expressions were obtained by combining the second term within
the expectation value appearing in (\ref{3.17}) and the second term on the
right-hand side of Eq. (\ref{3.21}). It should be taken into account that if
we put back the $\hbar $'s, there appears one with every noise kernel in
Eqs. (\ref{3.32}) and (\ref{3.33}).

The expressions (\ref{3.23}), (\ref{3.24}), (\ref{3.32}) and (\ref{3.33})
for $\delta \Omega (t)$, $A(t)$, $B(t)$ and $C(t)$, respectively, coincide
exactly with those of ref. \cite{halliwell96}, which are in turn equivalent
to those obtained in ref. \cite{hu92}. Thus, this derivation of the master
equation based on a stochastic description for the system is an alternative
to those given previously \cite{halliwell96,hu92,paz94} and is, of course,
in agreement with their results.

\section{DISCUSSION}

\label{sec5}

In this paper we have considered the stochastic description of a linear open
quantum system to give an alternative derivation of the corresponding master
equation. We have shown that the reduced Wigner function can be written as a
formal distribution function for a stochastic process characterized by a
Langevin-type equation. The master equation has then been deduced as the
corresponding Fokker-Planck equation for the stochastic process. This
derivation can be extended to the case of nonlinear interaction between
system and environment by computing perturbatively the influence action up
to quadratic order and even to the case of a general potential for the
system \cite{calzetta01b}.

It should be pointed out that whereas one can derive the Fokker-Planck
equation from the Langevin equation, the opposite is not possible in
general. One can always consider Langevin equations with stochastic sources
characterized by different noise kernels which, nevertheless, lead to the
same Fokker-Planck equation and, thus, the same master equation. This can be
argued from the expressions obtained in the derivation of the Fokker-Planck
equation. Let us consider, for simplicity, the situation corresponding to
local dissipation. A local contribution to the noise gives no contribution
to $B(t)$, but it does contribute to $C(t)$ as can be seen from Eqs. (\ref
{3.32}) and (\ref{3.33}) taking into account that $G_{ret}(t,t)=0$ and $%
\left. \partial G_{ret}(t^{\prime },t)/\partial t^{\prime }\right|
_{t^{\prime }=t}=M^{-1}$. Thus, one can always choose any noise kernel that
gives the desired $B(t)$ and then add the appropriate local contribution to
the noise kernel to get the desired $C(t)$ keeping $B(t)$ fixed. Note that
changing the noise kernel does not change $A(t)$. To illustrate the fact
that there exist different noise kernels giving the same $B(t)$, as was
stated above, one may consider the particular case corresponding to the weak
dissipation limit so that $G_{ret}(t,t^{\prime })\sim (M\Omega )^{-1}\sin
\Omega (t-t^{\prime })\theta (t-t^{\prime })$. To see that a different $%
\tilde{N}(t,t^{\prime })$ giving the same $B(t)$ as $N(t,t^{\prime })$
exists reduces then to show that there is at least one nontrivial function $%
\nu (s,t)=$ $\tilde{N}(t,t^{\prime })-N(t,t^{\prime })$ (with $s=t-t^{\prime
}$) such that for any $t$ $\int_{0}^{t}ds\sin (\Omega s)\nu (s,t)=0$, which
can be shown to be the case.

The fact that different Langevin equations lead to the same master equation
reflects that the former contains more information than the latter. To be
more precise, what we showed was that a Langevin equation contains in
general more information that the corresponding Fokker-Planck equation. To
extend this assertion to the master equation, one should make sure that
different Langevin equations leading to the same Fokker-Planck equation can
be obtained from an influence functional. Indeed this can be shown to be the
case provided that one considers general Gaussian initial states for the
environment. The inequivalence between the Langevin equation and the master
equation can be qualitatively understood in the following way. In the
influence functional it is only the evolution of the environment degrees of
freedom that is traced out. Of course, having integrated over all the
possible quantum histories for the environment, no correlations in the
environment can be obtained. Nevertheless, since the system is interacting
with the environment, non-Markovian correlations for the system at different
times may in general persist. On the other hand, when considering either the
reduced density matrix or its propagator, also the system evolution, except
for the final state, is integrated out. Consequently, information on
non-Markovian time correlations for the system is no longer available. Thus,
only when the system's reduced dynamics is Markovian, {\em i.e.} the
influence functional is local in time, we expect that the Langevin equation
and the master equation contain the same information. In particular, for a
Gaussian stochastic source, as in our case, the Langevin equation contains
the information about the system correlations at different times, which the
Fokker-Planck equation cannot in general account for. Only in the case in
which the dynamics generated by the Langevin equation is Markovian one can
compute the correlation functions just from the solutions of the
Fokker-Planck equation or, equivalently, the master equation for the
propagator $J(x_{2},x_{2}^{\prime },t_{2};x_{1},x_{1}^{\prime },t_{1})$; see
Eq. (\ref{2.6}). The key point is the fact that the propagator for the
reduced density matrix only factorizes when the influence action is local.
See ref. \cite{calzetta01} for a detailed argument on this point.

It is important to note that for a closed quantum system the evolution
determined by the time evolution operators $U(t_{2},t_{1})$ obtained from
the Schr\"{o}dinger equation is always unitary and, thus, also Markovian.
That is why the Schr\"{o}dinger equation suffices to get the correlation
functions for a closed quantum system. On the contrary, for an open quantum
system the evolution is nonunitary and, provided the influence action is
nonlocal, not even Markovian.

Finally, we should insist on the fact that, although we have exploited the
formal description of open quantum systems in terms of stochastic processes,
a classical statistical interpretation is not always possible. Thus,
although the Wigner function is a real and properly normalized function
providing a distribution for the initial conditions of our formal stochastic
processes, it is not a true probability distribution function in the sense
that it is not positive definite in general. In fact, this property is
crucial for the existence of quantum coherence for the system. Nevertheless,
even though the Langevin equation does not in general describe actual
classical trajectories of the system, it is still a very useful tool to
compute quantum correlation functions \cite{calzetta01} or even as an
intermediate step to derive the master equation.

\acknowledgements

We are grateful to Rosario Mart\'{\i }n for interesting discussions and to
Daniel Arteaga for a careful reading of the manuscript. This work has been
partially supported by the CICYT Research Project No. AEN98-0431 and by
Fundaci\'on Antorchas under grant A-13622/1-21. E.\ C.\ acknowledges support
from Universidad de Buenos Aires, CONICET and Fundaci\'on Antorchas.

\end{document}